\documentclass[transmag]{IEEEtran}
\usepackage[shortlabels]{enumitem}
\usepackage{graphicx}
\usepackage{epsfig}
\usepackage{amsmath,amsthm,amssymb}
\usepackage{float}
\usepackage{cases,setspace,adjustbox}
\usepackage{cite}
\usepackage{array}
\usepackage{eqparbox}
\usepackage{url}
\usepackage{xcolor}
\usepackage{mathtools}
\usepackage{bbm}
\usepackage{multirow}
\usepackage{multicol}

\usepackage{booktabs}
\usepackage{comment}
\usepackage{diagbox}
\usepackage[acronym]{glossaries}
\setacronymstyle{long-short}

\newacronym{AR}{AR}{Augmented Reality}
\newacronym{AI}{AI}{Artificial Intelligence}

\newacronym{BS}{BS}{Base Station}

\newacronym{CDF}{CDF}{cumulative distribution function}
\newacronym{CSI}{CSI}{Channel State Information}

\newacronym{D2D}{D2D}{Device-to-Device}
\newacronym{DC}{DC}{dual connectivity}
\newacronym{DSS}{DSS}{Dynamic Spectrum Sharing}

\newacronym{eNB}{eNB}{Evolved Node-B}
\newacronym{EPC}{EPC}{Evolved Packet Core}
\newacronym{eMTC}{eMTC}{enhanced MTC}

\newacronym{FDD}{FDD}{Frequency Domain Duplex}
\newacronym{FDM}{FDM}{Frequency Domain Multiplexing}

\newacronym{gNB}{gNB}{Next Generation Node-B}
\newacronym{Gbps}{Gbps}{gigabits per second}

\newacronym{HAP}{HAP}{High-Altitude Platform}

\newacronym{IEEE}{IEEE}{Institute of Electrical and Electronics Engineers}
\newacronym{IoE}{IoE}{Internet of Everything}
\newacronym{IoT}{IoT}{Internet of Things}
\newacronym{IIoT}{IIoT}{Industrial Internet of Things}
\newacronym{IoMT}{IoMT}{Internet of Medical Things}
\newacronym{IRS}{IRS}{Intelligent Reflecting Surfaces}
\newacronym{ITU}{ITU}{International Telecommunication Union}

\newacronym{KPI}{KPIs}{Key Performance Indicators}

\newacronym{LEO}{LEO}{Low Earth Orbit}
\newacronym{LTE}{LTE}{Long Term Evolution}
\newacronym{LTEAPRO}{LTE-A~Pro}{LTE Advanced Pro}

\newacronym{MCS}{MCS}{Modulation and Coding Scheme}
\newacronym{MM}{mmWave}{millimeter wave}
\newacronym{MIMO}{MIMO}{Multiple-Input Multiple-Output}
\newacronym{mMIMO}{mMIMO}{Massive Multiple-Input Multiple-Output}
\newacronym{mMTC}{mMTC}{massive machine-type communication}
\newacronym{MTC}{MTC}{Machine-Type Communication}
\newacronym{ML}{ML}{Machine Learning}
\newacronym{MU-MIMO}{MU-MIMO}{multi-user MIMO}
\newacronym{multi-TRP}{multi-TRP}{multiple transmission/reception point}

\newacronym{NR}{NR}{New Radio}
\newacronym{NTN}{NTN}{Non-Terrestrial Network}

\newacronym{PLS}{PLS}{Physical Layer Security}

\newacronym{QC}{QC}{Quantum Computing}
\newacronym{QoS}{QoS}{Quality of Service}

\newacronym{RAN}{RAN}{Radio Access Network}
\newacronym{RANs}{RANs}{Radio Access Networks}
\newacronym{RAT}{RAT}{Radio Access Technology}
\newacronym{RE}{RE}{Resource Element}
\newacronym{RHS}{RHS}{Reconfigurable Holographic Surface}
\newacronym{RIS}{RIS}{Reconfigurable Intelligent Surface}
\newacronym{RL}{RL}{Reinforcement Learning}
\newacronym{RU}{RU}{Radio Unit}
\newacronym{RT}{RT}{real-time}

\newacronym{TCI}{TCI}{Transmission Configuration Indicator}
\newacronym{TDD}{TDD}{Time Division Duplex}
\newacronym{TDM}{TDM}{Time Division Multiplexing}
\newacronym{TR}{TR}{Technical Report}
\newacronym{THz}{THz}{Terahertz}

\newacronym{UM-MIMO}{UM-MIMO}{ultra-massive MIMO}
\newacronym{URLLC}{URLLC}{ultra-reliable low-latency communication}
\newacronym{UAV}{UAVs}{Unmanned Aerial Vehicles}
\newacronym{UE}{UE}{User Equipment}
\newacronym{UL}{UL}{uplink}

\newacronym{V2V}{V2V}{Vehicle-to-Vehicle}
\newacronym{V2X}{V2X}{Vehicle-to-Everything}
\newacronym{VR}{VR}{Vitual Reality}

\newacronym{WLAN}{WLAN}{Wireless Local Area Network}
\newacronym{XR}{XR}{Extended Reality}
\newacronym{3GPP}{3GPP}{3rd Generation Partnership Project}
\newacronym{3D}{3D}{three-dimensional}

\newacronym{6GC}{6GC}{6G Core}
\newacronym{SBA}{SBA}{service-based architecture}
\newacronym{NF}{NF}{network function}
\newacronym{LLM}{LLM}{Large Language Model}
\newacronym{SBI}{SBI}{service-based interface}
\newacronym{RAG}{RAG}{Retrieval-Augmented Generation}
\newacronym{PCAP}{PCAP}{packet capture}
\newacronym{TS}{TS}{technical specification}
\newacronym{AF}{AF}{application function}
\newacronym{OAM}{OAM}{Operations, Administration, and Maintenance}
\newacronym{NWDAF}{NWDAF}{Network
Data Analytics Function}
\newacronym{AMF}{AMF}{Access and Mobility Management Function}
\newacronym{SMF}{SMF}{Session Management Function}
\newacronym{UPF}{UPF}{User Plane Function}
\newacronym{GUI}{GUI}{graphical user interface}
\newacronym{CLI}{CLI}{command-line interface}
\newacronym{NRF}{NRF}{NF Repository Function}
\newacronym{LNF}{LNF}{LLM-based Network Function}
\newacronym{RESTful}{RESTful}{REpresentational State Transfer}
\newacronym{NEF}{NEF}{Network Exposure Function}
\newacronym{LAF}{LAF}{LLM-based Application Function}
\newacronym{DN}{DN}{data network}
\newacronym{SLA}{SLA}{Service Level Agreement}
\newacronym{MNO}{MNO}{Mobile Network Operator}
\newacronym{NOC}{NOC}{Network Operations Center}
\newacronym{PFCP}{PFCP}{Packet Forwarding Control Protocol}
\newacronym{NGAP}{NGAP}{ Next Generation Application Protocol}
\newacronym{IVR}{IVR}{interactive voice response}
\newacronym{FAQ}{FAQ}{frequently asked question}
\newacronym{eMBB}{eMBB}{enhanced mobile broadband}
\newacronym{GBR}{GBR}{Guaranteed Bit Rate}
\newacronym{ARP}{ARP}{Allocation and Retention Priority}
\newacronym{PCF}{PCF}{Policy Control Function}
\newacronym{PDU}{PDU}{protocol data unit}
\newacronym{BERT}{BERT}{Bidirectional Encoder
Representations from Transformers}
\newacronym{SOTA}{SOTA}{state-of-the-art}
\newacronym{SON}{SON}{self-organizing network}
\newacronym{GPT}{GPT}{Generative Pre-trained Transformer}
\newacronym{LLaMA}{LLaMA}{Large Language Model Meta AI}
\newacronym{GPU}{GPU}{Graphics Processing Unit}
\newacronym{CoT}{CoT}{Chain-of-Thought}
\newacronym{ToT}{ToT}{Tree of Thoughts}
\newacronym{Near-RT}{Near-RT}{near-real-time}
\newacronym{Non-RT}{Non-RT}{none-real-time}
\newacronym{IETF}{IETF}{Internet Engineering Task Force}
\newacronym{AISV}{AISV}{AI independent software vendor}
\newacronym{SbD}{SbD}{security-by-design}

\newacronym{IC}{IC}{Immersive Communication}
\newacronym{HRLLC}{HRLLC}{Hyper Reliable and Low Latency Communication}
\newacronym{MC}{MC}{Massive Communication}

\makeatletter
 \let\old@ps@headings\ps@headings
 \let\old@ps@IEEEtitlepagestyle\ps@IEEEtitlepagestyle
 \def\confheader#1{%

 \def\ps@IEEEtitlepagestyle{%
 \old@ps@IEEEtitlepagestyle%
 \def\@oddhead{\strut\hfill#1\hfill\strut}%
 \def\@evenhead{\strut\hfill#1\hfill\strut}%
 }%
 \ps@headings%
 }
 \makeatother

\confheader{%
\begin{minipage}{\textwidth}
\centering \small
This paper has been submitted to IEEE for publication. Copyright may change without notice.
\end{minipage}
 }

\begin{document}
\title{Mobile Network-specialized Large Language Models for 6G: Architectures, Innovations, Challenges, and Future Trends}
\author{{{Abdelaali Chaoub},~\IEEEmembership{Senior Member, IEEE},
{Muslim Elkotob},~\IEEEmembership{Member, IEEE}
}

\thanks{
Abdelaali Chaoub is with the Institut National des Postes et Télécommunications (INPT), Morocco (email: chaoub.abdelaali@gmail.com).
Muslim Elkotob is with Vodafone, Germany (email: muslim.elkotob@vodafone.com).
}
}
\maketitle
\section*{Abstract}
\label{sec:abstract}
Conventional 5G network management mechanisms, that operate in isolated silos across different network segments, will experience significant limitations in handling the unprecedented hyper-complexity and massive scale of the sixth generation (6G). Holistic intelligence and end-to-end automation are, thus, positioned as key enablers of forthcoming 6G networks. The Large Language Model (LLM) technology, a major breakthrough in the Generative Artificial Intelligence (AI) field, enjoys robust human-like language processing, advanced contextual reasoning and multi-modal capabilities. These features foster a holistic understanding of network behavior and an autonomous decision-making. This paper investigates four possible architectural designs for integrated LLM and 6G networks, detailing the inherent technical intricacies, the merits and the limitations of each design. As an internal functional building block of future 6G networks, the LLM will natively benefit from their improved design-driven security policies from the early design and specification stages. An illustrative scenario of slicing conflicts is used to prove the effectiveness of our architectural framework in autonomously dealing with complicated network anomalies. We finally conclude the paper with an overview of the key challenges and the relevant research trends for enabling Mobile Network-specialized LLMs. This study is intended to provide Mobile Network Operators (MNOs) with a comprehensive guidance in their paths towards embracing the LLM technology.

\section{Introduction}
\label{sec:intro}
The upcoming 6G networks are widely anticipated to achieve unprecedented levels of autonomous operation and self-X functionalities \cite{chaoub2023hybrid}. These expectations are primarily driven by the recent trend toward seamless integration of \gls{AI} capabilities in the telecommunication landscape. In the meantime, \gls{AI} development has witnessed a remarkable progress, over the last decade, largely due to the cutting-edge transformer schemes \cite{vaswani2017attention} and the explosion of \glspl{LLM}. Both technologies substantially revolutionized natural language understanding- and generation-related tasks, and demonstrated a great potential to drive innovation in various industries.

The alliance between 6G networks and \glspl{LLM} heavily relies on \gls{LLM} access to rich and specialized knowledge stemming from the accumulated experience of various telecom stakeholders, in particular, \glspl{MNO}. Nowadays, there is a growing interest in adapting various \gls{SOTA} \glspl{LLM} to the telecommunication ecosystem \cite{maatouk2024large}. Examples of \gls{LLM}-empowered \gls{MNO}'s scenarios include network analysis (e.g. using \gls{LLaMA} \cite{Kan2024mobile_llama}), network configuration (e.g. proof-of-concept utilizing \gls{GPT} 4 in \cite{Wang2024NetConfEval}), fault diagnosis (e.g. based on \gls{BERT} \cite{Chen2023Knowledge}), 
standards documentation understanding (e.g. comparative assessment among fine-tuned \gls{LLaMA}3, Solar and Mistral against \gls{GPT}-4 in \cite{Said2024instruct}), and \glspl{SON} \cite{Bariah2024Large}. A comprehensive survey is provided in \cite{Zhou2024survey} for broader details. Nevertheless, it can be clearly seen that the literature on this topic is predominantly focusing on the potential use cases and the resulting gains from integrating the \glspl{LLM} in the realm of 6G networks, leaving the details of how this integration can be implemented in real-world settings largely unexamined. To the best of our knowledge, this paper is the first attempt to provide an in-depth analysis of the architectural design possibilities for an integrated \gls{LLM} and 6G technology. We advocate for incorporating the \gls{LLM} as an intrinsic building block of the \gls{MNO} infrastructure synergistically operating with the remaining components, rather than operating solely as a supplementary add-on or on-demand service. This deep integration will particularly benefit from the \gls{SbD} approach adopted by 6G \cite{Khaloopour2024Resilience}, wherein security is embedded at every stage of the network’s design, deployment, and operation.

Based on the above introduction, the contributions of this article are two-fold:
\begin{itemize}
    \item We build upon the work in \cite{Kan2024mobile_llama}, wherein the focus is on augmenting the \gls{NWDAF} with \gls{LLM} capabilities. \gls{NWDAF} is a \gls{3GPP} network function that collects data from various Core \glspl{NF}, performs network statistical and predictive analytics to be shared with authorized data consumers. We extend the analysis by investigating alternative architectures for incorporating the \gls{LLM} component as a part of 6G \glspl{NF} while describing the merits and the shortcomings of each possible solution. This offers a comparative assessment that assists mobile carriers in understanding the trade-offs and applicability of different architectures. We additionally outline the architectural innovations required to implement these designs.
    \item We provide a concrete example of a complicated anomalous situation to illustrate the inter-exchanges that the \gls{LLM} triggers with the rest of the network to autonomously remedy this situation. 
    \end{itemize}

The remainder of this article is organized as follows. In the next section, we describe how the \gls{MNO} ecosystem can benefit from \glspl{LLM} through features like fine-tuning and prompt engineering. Afterwards, we introduce the possible architectures for building \gls{LLM}-enabled 6G networks, emphasizing the pros and cons of each option. Later, we detail different \gls{LLM} call flows allowing an autonomous resolution of a concrete faulty scenario in 6G networks. The associated challenges and directions of research are identified in the following section. Lastly, we conclude the article.
\section{Mobile Networks Powered by \glspl{LLM}}
\label{sec:LLM_enabled_MNs}
\begin{figure*}[t!]
\centering
\includegraphics[width=.99\textwidth]{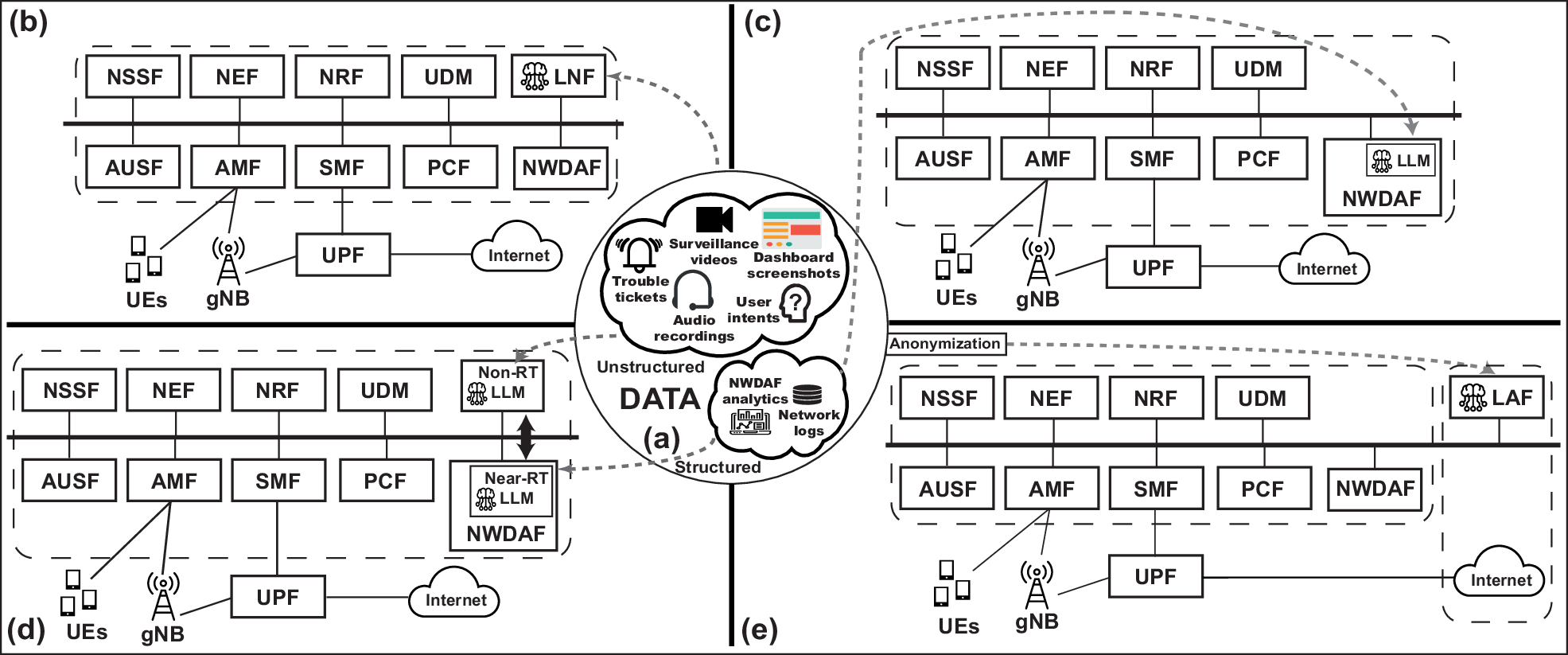}
    \caption{Possible architectural designs for integrated \gls{LLM} and \gls{MNO} ecosystem.}
    \label{fig:LLM_possible_architectures}
\end{figure*}
The historical data of the \gls{MNO}, archived over years of expertise, constitutes a solid foundation for training the \gls{LLM} using structured and unstructured multi-modal inputs (as illustrated in Fig.~\ref{fig:LLM_possible_architectures}a) such as user intents, network logs, alarm descriptions, trouble tickets, \gls{PCAP} files (e.g. from Wireshark or tcpdump), dashboard screenshots, audio recordings (e.g. from \gls{IVR} systems), video feeds (e.g. from infrastructure surveillance), and \gls{NWDAF} analytics. To this end, a separate collection framework aggregates data from various sources into a centralized repository, and extracts most informative features such as warnings, error codes, timestamps, and user/gNB/session/bearer/\gls{QoS} flow/slice IDs. The extracted features are then converted into unified embeddings that are combined into a common vector space with suitable metadata (e.g. to differentiate data formats). The resulting vector store is used to fine-tune the \gls{LLM} to deeply internalize \gls{MNO}-specific knowledge \cite{Bariah2023understanding}. This allows the \gls{LLM} to learn patterns, sequences, and deviations that correlate with normal or faulty network operations. This is made possible using a timestamp-based cross-referencing to link different entries from several data sources, allowing detailed description and context for each flagged event as well as the resolution workflow for the spotted anomalies.

In live mobile networks, fresh multi-modal data is continuously fed into the \gls{LLM}, either uploaded in batches or streamed in real-time. The \gls{LLM} analyzes this data and identifies potential anomalous behaviors in light of its accumulated learning. In case of new anomalies not covered during the fine-tuning stage, the \gls{LLM} can rely on clustering techniques to group similar patterns and flag outliers as suspected behaviors. The \gls{LLM} is also capable of using \gls{RAG}-enabled external knowledge databases such as \gls{3GPP} documents \cite{Said2024instruct}, \gls{IEEE} standards, \gls{IETF} RFCs and vendors documentation \cite{soman2023observations} to compare the actual network behavior with the expected one to identify misconfigurations and spot unusual trends in protocols and communication flows. Well-crafted prompts, on the other hand, can guide the \gls{LLM} responses to provide focused solutions. Paradigms such as the \gls{CoT} reasoning can be used to break down the \gls{LLM} insights into a series of simplified and actionable sub-tasks. It can be extended by the \gls{ToT} technique to explore different reasoning paths and identify the most optimal solution. The \gls{LLM} can naturally produce stepwise reasoning if datasets used for fine-tuning contain \gls{CoT} and \gls{ToT} examples, or through creative prompting \cite{Zhou2024survey}. In parallel, \gls{NOC} engineers can intervene to confirm, guide or reject the \gls{LLM} findings, if needed, e.g. using its intuitive conversational interface. Through continuous self-learning, the \gls{LLM} will dynamically adapt to evolving network conditions, optimizing its performance over time \cite{Chaparadza2023optimization}.


By incorporating \glspl{LLM} (e.g. as \glspl{NF}) into upcoming 6G networks, expected to be designed with \gls{SbD} principles \cite{Khaloopour2024Resilience}, \glspl{LLM} will naturally inherit the same built-in security safeguards rather than adding them as an afterthought. This design-driven approach focuses on proactive threat management, end-to-end encryption, authentication, network slicing isolation, \gls{AI}-driven threat detection with automated reactions, and stateless designs, fostering a resilient \gls{LLM}.

\section{Deployment Strategies of \glspl{LLM} in Mobile Networks}
\label{sec:LLM_deployments}
Below, we describe the key forms an \gls{LLM}-empowered 6G network can take with a thorough review of the advantages and drawbacks of each approach:

\subsection{\gls{LLM} as a standalone \gls{NF} (see Fig.~\ref{fig:LLM_possible_architectures}b)}
\label{subsec:standalone_llm}
\begin{figure*}[t!]
\centering
\includegraphics[width=.99\textwidth]{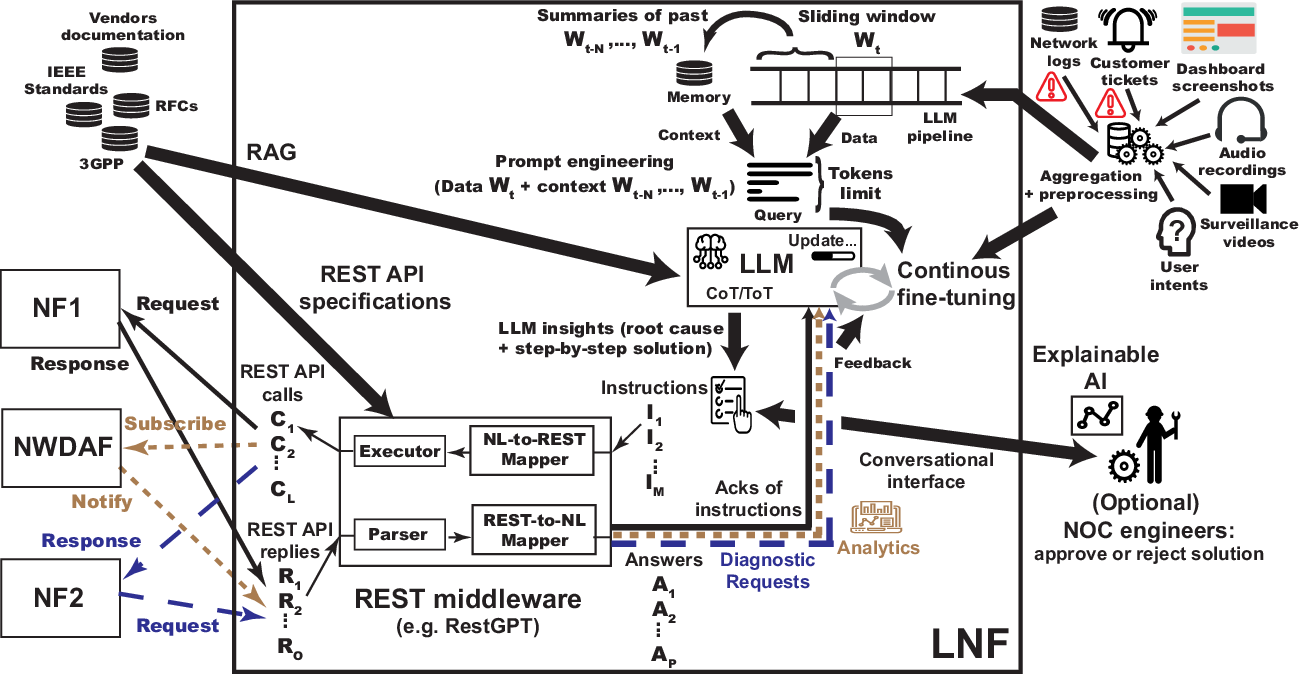}
    \caption{A breakdown of the \gls{LNF}.}
    \label{fig:LNF_breakdown}
\end{figure*}

The \gls{LNF} will be deployed as an independent \gls{NF} at the control plane of the \gls{6GC} as a part of the \gls{SBA}. It will interact with other \glspl{NF} via the \glspl{SBI} using stateless \gls{RESTful} APIs that are based on HTTP/2 and JSON payloads. These interactions typically follow the synchronous request/response or the asynchronous subscribe/notify models, depending on the use case. For instance, an \gls{NF} can request a diagnostic from the \gls{LNF} about an incident, confirms the initiation of a resolution workflow or subscribes to be notified about anomalies in network behavior. In turn, the \gls{LNF} can request the execution of a step-by-step healing plan upon detection of an anomaly, or subscribes to specific network analytics, as shown in Fig.~\ref{fig:LNF_breakdown}. The \gls{LNF} needs to first register its services with the \gls{NRF} to be discoverable by the remaining \glspl{NF}.

A breakdown of the \gls{LNF} is provided in Fig.~\ref{fig:LNF_breakdown}. First, the \gls{MNO}'s multi-modal and time-evolving data should be be fed to the processing pipeline of the \gls{LLM}. Afterwards, relevant information should be identified, extracted and structured according to a consistent format. We use a periodic sliding window mechanism that controls how much data, $W$, can be sent to the \gls{LLM} prompt per fixed-time period, $T$. At the end of each period, the window moves forward and the next data is transmitted to the \gls{LLM} in the subsequent time period. The preprocessed data is then incrementally submitted to the \gls{LLM} at regular time intervals. Meanwhile, the \gls{LLM} leverages its natural language capabilities to generate a compact summary of past key information (e.g. metrics, events, and network changes) from the preceding time windows, to be used as a context for the present sliding window $W_{t}$. The context should be concise yet detailed enough to capture significant patterns in the past. For example, the preprocessed data can be summarized every 1-hour interval and then the most recent $N$ summaries can be provided to the \gls{LLM} as a context, or in case of long-term anomalies, fine-grained summaries can be provided for recent data and coarser summaries can be generated for older data. This calls for a long-term memory to retain information over extended intervals. The size $W$ and periodicity $T$ of sliding window should be adaptively chosen to balance enough context for accurate predictions and maintaining the overall input (recent data+context) within token limits. Next, prompt engineering techniques are used to craft a natural language query that includes the most recent sliding window data, $W_{t}$, along with summaries of the past $N$ iterations $\{W_{t-N}, \ldots, W_{t-1}\}$ as a context, to be submitted to the \gls{LLM}. Upon detecting an anomaly, the \gls{LLM} will provide its insights including the diagnosis (i.e. root cause) and the suggested solution. Using \gls{CoT} prompting, the \gls{LLM} decomposes the solution into a series of $M$ sequential and manageable instructions $\{I_{1}, \ldots, I_{M}\}$. Further, \gls{ToT} prompting can enhance the clarity and readability of the generated instructions. Thereafter, each instruction, or set of instructions, is mapped into a single \gls{3GPP} \gls{RESTful} API calling structure using an NL-to-REST Mapper. The latter consults the appropriate \gls{3GPP} \gls{TS}, that describes the endpoints, operations, request/response formats, parameters, error handling, authentication, and other details, to generate the correct API calls, $\{C_{1}, \ldots, C_{L}\}$. Later, an Executor sub-module invokes each resulting API call using an URL pointing towards the relevant \gls{NF}  to run the instructions at hand. The resulting execution outputs $\{R_{1}, \ldots, R_{O}\}$, in particular the JSON payload, will be parsed according to the \gls{3GPP} API documentation to extract useful information using a Parser sub-module. Then, the relevant information will be presented in an actionable format, $\{A_{1}, \ldots, A_{P}\}$, to acknowledge that the step-by-step resolution plan advised by the \gls{LLM} was successfully triggered within the network. The REST-to-NL Mapper is used to translate \gls{3GPP} \gls{RESTful} API replies into natural language answers. RestGPT \cite{song2023restgpt} represents a good attempt to develop a framework connecting \glspl{LLM} with \gls{RESTful} APIs. The \gls{LLM} updates its fine-tuning knowledge through a continuous improvement cycle using different query-feedback pairs.

This modular design ensures that \gls{LLM} operations enjoy dedicated compute and storage resources and the \gls{LNF} can scale independently from the remaining \glspl{NF} as workload grows. Moreover, any updates, maintenance or replacements to the \gls{LLM} or the \glspl{NF} can be performed independently without or with minimal disruption. The \gls{LNF} can also interface with multiple network entities, providing multi-functional support across other \glspl{NF} (e.g. \gls{SMF}) and external business systems (e.g. Metaverse, automotive, and quantum computing). The integration within the \gls{SBA} ensures stateless communications (i.e. REST architecture), slicing-based fault isolation, full encryption (e.g. using TLS) and restricted access to legitimate services (e.g. using OAuth), reducing attack surfaces \cite{Khaloopour2024Resilience}.

Nevertheless, additional middleware is needed for the \gls{LLM}, as depicted in Fig~\ref{fig:LNF_breakdown}, to interact with the remaining network components leading to potential delays and complexities in communication. This impediment is further exacerbated by the need to setup real-time feedback loops, at the intelligence level, between the \gls{LNF} and the remaining \glspl{NF} especially the \gls{NWDAF} to synchronize \gls{AI}-related insights and avoid conflicts and duplication.

\subsection{LLM as an embedded intelligence layer (see Fig.~\ref{fig:LLM_possible_architectures}c)}
\label{subsec:integrated_llm}
6G \glspl{NF} can be augmented with \gls{LLM} capabilities, endowing them with the ability to perform innovative functions such as advanced predictive analytics and user-/operator-friendly interactions. The \gls{LLM} can be centrally hosted within a single \gls{NF} (e.g. \gls{LLM}-augmented \gls{NWDAF} in \cite{Kan2024mobile_llama}) or distributed across multiple \glspl{NF}.

A major advantage of \gls{LLM}-powered \glspl{NF} is improved interaction between the \gls{NOC} engineers and the 6G ecosystem, using conversational queries instead of complex \glspl{CLI}. Moreover, embedding the \gls{LLM} into an \gls{NF} eliminates the need for intermediary communication interfaces, simplifying the overall system flows. The \gls{LLM} will also gain direct and real-time access to the established pipeline of the \gls{NF} to avoid redundant computations and conflicting outputs. For example, the analytics function will be seamlessly coordinated within the \gls{LLM}-augmented \gls{NWDAF}  \cite{Kan2024mobile_llama}.

On the downside, this fully integrated approach may require significant architectural changes at the \gls{NF} level, e.g. to embed large-scale neural networks. Further, the \gls{LLM} scalability will be constrained by the hosting \gls{NF}'s design, as they will compete for resources. While conventional \glspl{NF} (e.g. \gls{AMF}, \gls{UPF}) scale primarily to handle higher traffic volumes and user connections, an \gls{LLM} may scale for different \gls{AI}-specific considerations such as \gls{LLM} upgrades to update model architecture. On the other hand, the \gls{LLM} functionalities become tightly coupled with the hosting \gls{NF}, limiting the \gls{LLM}'s applicability outside the \gls{NF}'s primary goal (e.g. the \gls{LLM}-augmented \gls{NWDAF} design \cite{Kan2024mobile_llama} will restrict the \gls{LLM} capacity to provide novel services beyond the analytics function of the \gls{NWDAF}). Further, a failure in the integrated design can affect both the \gls{LLM} and the \gls{NF}.

\subsection{Hybrid Approach: partially integrated \gls{LLM} closely coupled with an existing \gls{NF} (see Fig.~\ref{fig:LLM_possible_architectures}d)}
\label{subsec:hybrid_llm}
The hybrid approach involves integrating a lightweight \gls{LLM} sub-component within an existing \gls{NF} (e.g. \gls{NWDAF}) and deploying a more sophisticated \gls{LLM} sub-component outside the \gls{NF}. This approach strategically integrates the basic short-term features of \gls{LLM} such as structured data-based analytics into the hosting \gls{NF}, within a \gls{Near-RT} \gls{LLM} module, whereas the breakthrough long-term functions such as context-aware decision-making that rely on unstructured data are deployed externally as an independent microservice or containerized application in a \gls{Non-RT} \gls{LLM} \gls{NF}. Both modules closely communicate with each other using the established \gls{SBI}, thus building a loosely coupled design \cite{chaoub2023hybrid} that reduces the dependencies and constraints that one component places on another while maintaining necessary interactivity. The internal \gls{LLM} investigates the reported issues and categorizes them into \gls{Near-RT} and \gls{Non-RT}. The hosting \gls{NF} handles \gls{Near-RT} incidents immediately impacting network operations. It potentially requests additional analysis from the external \gls{LLM} module regarding long-term tasks, the latter then complements by providing deeper \gls{Non-RT} insights or endorses \gls{Near-RT} findings, sending them back to the hosting \gls{NF} or to the rest of the network for dissemination. The hybrid architecture potentially needs to incorporate an additional faster API (e.g. gRPC) to internally synchronize the analysis and the derived insights between \gls{Near-RT} and \gls{Non-RT} parts of \gls{LLM} to avoid slowing down the \gls{SBI}.

The hybrid design ensures that any updates or replacements to the \gls{Non-RT} \gls{LLM} or the hosting \gls{NF} can be performed with minimal disruption. The hosting \gls{NF} remains operational even if the external \gls{LLM} becomes unavailable, ensuring continuity of advanced statistical and predictive analytics. Furthermore, the strategic and the lightweight parts of the \gls{LLM} can scale independently. Yet another benefit is retaining the \gls{Non-RT} \gls{LLM} ability to independently connect to other \glspl{NF} or external stakeholders.

This architecture faces some technical hurdles related to its increased complexity and costs, and the need to carefully coordinate the \gls{Near-RT} and the \gls{Non-RT} components of the \gls{LLM} through the internal fast interface. The operational scope of each component should be clearly defined.
\subsection{\gls{LLM} as a standalone external \gls{AF} (see Fig.~\ref{fig:LLM_possible_architectures}e)}
\label{subsec:external_llm}
In this scheme, the \gls{LLM} is deployed as an \gls{AF}, referred to as \gls{LAF}, and resides outside the \gls{MNO}'s infrastructure. The \gls{LAF} communicates with the \gls{MNO} entities via \gls{RESTful} APIs, leveraging \gls{3GPP} standard interfaces defined for \glspl{AF}. As an external entity, the \gls{LAF} can be integrated with the \gls{SBA} via the \gls{NEF}, acting as a secure gateway for exposing network capabilities to external functions. As an alternative, the \gls{LAF} can be deployed inside a trusted \gls{DN}, a secure extension of the network, to directly communicate with relevant \gls{6GC} functions.

A primary strength of this scheme lies in the capacity of upgrading, scaling, or integrating the \gls{LAF} with external platforms independently of the \gls{6GC} architecture. External computing infrastructures, in particular, can be used to offload resource-intensive \gls{LLM} tasks to alleviate the burden on network resources. At the same time, the \gls{LAF} can leverage data from external domains (e.g., third-party services, public datasets) to enhance its insights. Another compelling advantage is allowing \glspl{MNO} to collaborate with \glspl{AISV}, leveraging best-in-class players and their expertise in \gls{LLM} solutions.

An externalized \gls{LLM} may, however, lead to an increased latency in real-time communications delaying the resolution of critical network outages. Beyond this, the \gls{LAF} as an external entity may lack complete real-time contextual awareness about network conditions, degrading the accuracy of its outcomes. In addition, exposing sensitive network and user data to the external \gls{LAF} requires robust encryption, strict access control, anonymization, and full compliance with regulations in terms of data protection, privacy and sovereignty. The \gls{NEF}, in particular, need to be secure enough to handle different exchanges between the local \glspl{NF} and the external \gls{LAF} to avoid any security breaches. One further impediment arises from the strong dependency on \glspl{AISV} regarding service reliability and availability. Lastly, the ability to customize \gls{LLM} services to specific needs of the \gls{MNO} may be limited.

\begin{table*}[t!]
\centering
\footnotesize
\setlength{\belowcaptionskip}{0.33cm}
\caption{Potential strategies to integrating \glspl{LLM} within 6G networks: strengths and weaknesses.}
\label{tab:architectures_table}
\renewcommand{\arraystretch}{1.5}
\begin{tabular}{|m{3.5cm}|m{6.5cm}|m{6.5cm}|}
\hline
\textbf{Architecture}  & \textbf{Strengths} & \textbf{Weaknesses} \\
\hline
\multirow{5}{3.5cm}{Standalone \gls{LLM} (Subsec.~\ref{subsec:standalone_llm})} & Independent and extended operational range of the \gls{LLM} with a wider spectrum of usecases & Increased latency as data need to be routed through multiple internal entities/interfaces to reach the \gls{LLM} \\
& Interoperability with multiple entities and systems both internally and externally & The need for additional middleware to interface with the other network entities \\ 
& Dedicated resources &	Duplication of functionalities due to silos of intelligence deployed across the \glspl{NF}\\ 
& Upgrades and maintenance operations can be carried out without any cross-effect between the \gls{LLM} and the \glspl{NF} & 	\\
& Scalability and adaptability to dynamic workloads & 	\\
& \gls{SbD} inherited from early 6G efforts & 	\\
\hline

\multirow{6}{3.5cm}{Fully integrated \gls{LLM} (Subsec.~\ref{subsec:integrated_llm})} & Streamlined holistic design avoiding the need for intermediary communication middleware & Complex integration due to the required architectural developments at both cohabiting frameworks \\
& Shared access to the already established \gls{NF} pipeline and interfaces with outside &  Competition for the finite resources pool leading to resource contention\\ 
& Real-time insights as a result of seamless coordination with the \gls{NF} & Limited scalability due to shared resources and conflicting scaling factors \\ 
& Improved interaction between network operators and the \gls{NF} using conversational interfaces & Strong coupling and dependence on the \gls{NF} limiting the operational scope of the \gls{LLM}\\ 
& Enhanced contextual awareness at the \gls{NF} level & Risk of monolithic failure \\ 
& \gls{SbD} inherited from early 6G efforts & Interactions restricted to network entities or external business systems already interfacing with the \gls{NF}\\
\hline

\multirow{5}{3.5cm}{\begin{tabular}[c]{@{}l@{}}Partially integrated \gls{LLM} \\ (Subsec.~\ref{subsec:hybrid_llm}) \end{tabular}} & Balances modularity and integration & Complex intra-\gls{NF} signaling\\ 
& Decoupled scalability allowing for external scaling of the advanced \gls{LLM} capabilities & Development and maintenance overhead of both internal and external \gls{LLM} parts \\
& Upgrades and maintenance operations can be carried out with minimal cross-effect between the \gls{LLM} and the \gls{NF} & Synchronization overhead between the external part of the \gls{LLM} and the \gls{NF} \\ 
&  Loose coupling due to separate and decoupled functionalities between the external \gls{LLM} and the \gls{NF} & Complex architecture and costly implementation	\\ 
& Near real-time responsiveness since data flows through less entities and as a result of the fast internal interface & \\
& Dedicated resources &	\\ 
& \gls{SbD} inherited from early 6G efforts & 	\\
\hline

\multirow{5}{3.5cm}{\begin{tabular}[c]{@{}l@{}}External \gls{LLM} \\ (Subsec.~\ref{subsec:external_llm}) \end{tabular}} & Can be scaled independently of the \gls{MNO}'s architecture & The need to comply with data sovereignty laws and strengthen security measures to avoid sensitive data leaks\\ 
& Upgrades and maintenance operations can be carried out without any cross-effect between the \gls{LLM} and the network & Increased latency as data need to pass through multiple internal and external entities/interfaces to reach the \gls{LLM}\\ 
& Possibility to leverage data from external domains for enhanced insights & Lack of contextually relevant information about the internal status of the system\\ 
& Close collaboration between \glspl{MNO} and \glspl{AISV} for the latest best-in-class \gls{LLM} innovations & Less flexibility in terms of providing customized \gls{LLM} services meeting the \gls{MNO}'s specific needs\\ 
& Harnessing external powerful computing platforms to handle intensive \gls{LLM} processing & Dependency on third parties with regards to service reliability, availability and performance\\ 
\hline
\end{tabular}
\end{table*}

A comprehensive summary of the strengths and weaknesses of each approach is provided in Table~\ref{tab:architectures_table}.

\section{A Practical Example Showcasing the Autonomous \gls{LLM} Operations in Future 6G Networks}
\label{sec:LLM_applications}
In this section, we delve into a scenario that involves a resource contention between three slices within a 6G network. A first slice, $I$, devoted to \glspl{IC} such as holographic live streaming. A second slice, $H$, assigned to \glspl{HRLLC} including industrial \gls{IoT} applications. A last slice, $M$, dedicated to \glspl{MC} such as \gls{IoT} sensors. The predefined bandwidth allocation among the three slices is 50\%, 30\%, and 20\%, respectively. The latency upper bound for slice $H$ is 1 ms. We consider typical \gls{QoS} policy rules prioritizing \gls{IC} traffics over non-critical \gls{IoT} but ensuring that the stringent reliability requirements of \gls{HRLLC} are never violated. Due to an unforeseen large-scale event (e.g. a concert), slice $I$ experiences a surge in traffic, up to 80\% utilization, consuming excessive compute and bandwidth resources in the \gls{UPF}. This led to an increased latency up to 2 ms in slice $H$, and left insufficient capacity to efficiently meet its bandwidth needs.

In conventional 5G networks, the \gls{NWDAF} continuously monitors the performance of each slice, and spots the slice $I$'s resource overuse as well as the latency spikes for slice $H$. Thereby, the \gls{PCF} is notified to perform appropriate policy-based corrective actions. The \gls{PCF} generates \gls{QoS} updates to deprioritize slice $I$ traffic, e.g. through reducing the allocated bandwidth from 50\% to 30\% and altering the inherent \gls{ARP}, to free up resources for slice $H$. Policy updates are shared with the \gls{SMF} to adjust the corresponding \gls{QoS} parameters. The \gls{SMF} mandates the \gls{AMF} to enforce these parameters via modifying \gls{QoS} rules in the affected \gls{PDU} sessions and adjusting bearer priorities for slice $I$ over the impacted gNBs and UEs. Finally, \gls{QoS} updates are pushed to the \gls{UPF} via \gls{PFCP}, e.g. to reallocate more bandwidth towards slice $H$ and drop non-critical slice $I$ flows. Nevertheless, human decision-making and manual intervention is mandatory to extend the predefined \gls{PCF} policy rules. The \gls{MNO}, in consultation with the external service providers (i.e. the holographic communications platform and the industrial \gls{IoT} service provider) or the corresponding \glspl{SLA}, decided that the least disruptive strategy to restore slice $H$'s \gls{SLA} compliance is to reduce slice $I$'s allocated bandwidth from 50\% to 30\%. This, unfortunately, extends the disruption time.

\begin{figure*}[t!]
\centering
\includegraphics[width=.99\textwidth]{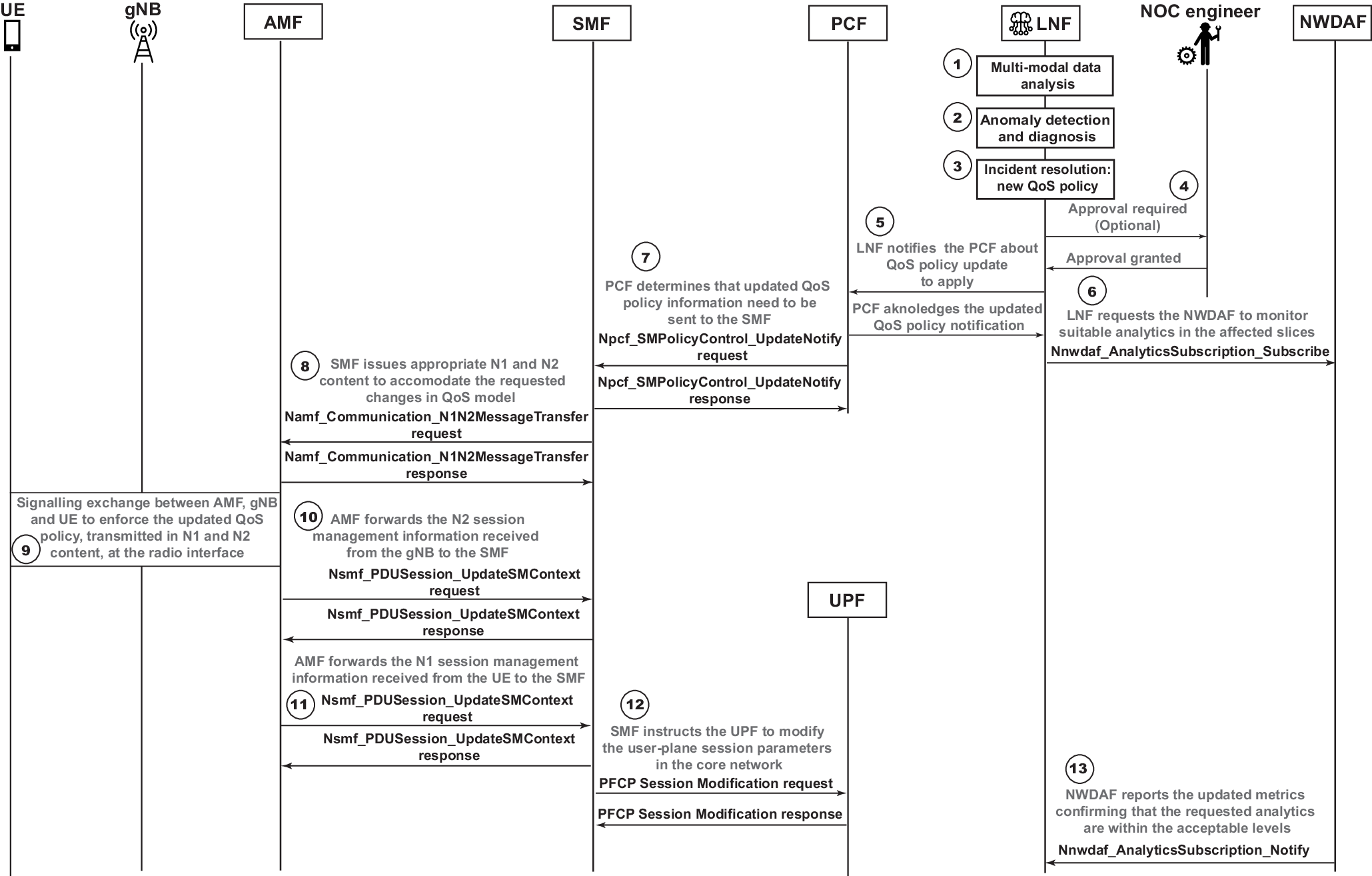}
    \caption{Workflow of \gls{LNF} autonomous operation for anomaly detection, diagnosis and resolution.}
    \label{fig:LNF_exchanges_with_NFs}
\end{figure*}

The \gls{LNF}-based architecture can autonomously and efficiently handle such slicing conflicts. As disparate data get processed by the \gls{LNF}, it tracks the traffic shape of each slice and will capture an unusual pattern indicative of a potential traffic spike in slice $I$. The fine-tuned and \gls{RAG}-powered \gls{LNF} conducts in-depth network-level analysis, identifies the root-cause (i.e. unexpected major event), alerts \gls{NOC} engineers and triggers proactive slicing adjustments to anticipate the anomaly occurrence. When the \gls{LNF} fails to proactively identify the unusual traffic pattern, it can still react to the traffic spike and the latency increase as they occur in real time. Based on its reasoning and optimization capabilities, the \gls{LNF} can find the optimal and precise \gls{QoS} policy that resolves this conflict, e.g. multi-objective optimization among the conflicting slices identifies the Pareto-optimal solutions and selects the most optimal bandwidth allocation for slice $I$, subject to \gls{MNO}'s approval. The \gls{LNF}'s explainable \gls{AI} will reinforce the \gls{MNO} confidence in the \gls{LLM} reasoning, ensuring a fast decision. Then, the required \gls{QoS} policy modifications will be communicated towards the \gls{PCF} for immediate consideration, and will be enforced over the involved \glspl{NF} following the same workflow as the conventional case. Meanwhile, the \gls{LNF} requests the \gls{NWDAF} to monitor throughput- and latency-related metrics in the affected slices. The \gls{NWDAF} reports that both metrics are within their acceptable intervals and thus acknowledging the effectiveness of the enforced actions. The entire resolution workflow is illustrated in Fig.~\ref{fig:LNF_exchanges_with_NFs}. The \gls{LNF} can also generate a report documenting the diagnosis, actions taken, and insights gained for \gls{MNO} records.

\section{Challenges, Trends, and Research Directions}
\label{sec:Challenges_and_future_outlook}
\begin{figure}[t!]
\centering
\includegraphics[width=.4\textwidth]{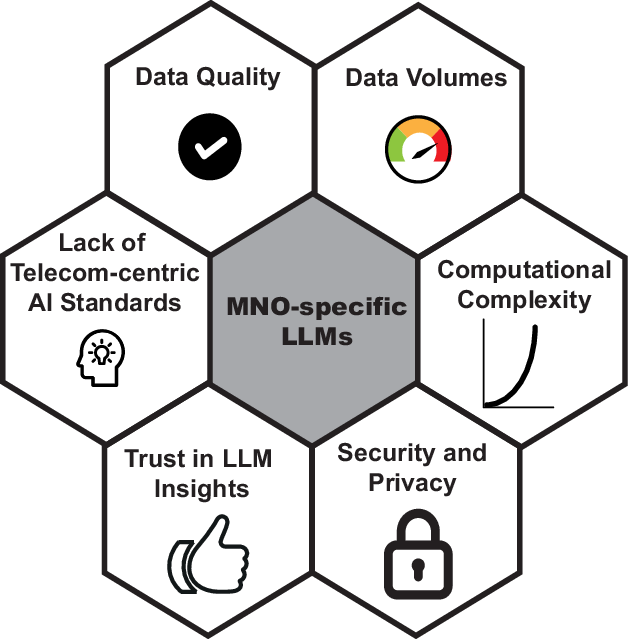}
    \caption{Key challenges for \gls{LLM}-powered 6G.}
    \label{fig:LNF_challenges}
\end{figure}

This section highlights the fundamental
challenges that may hinder the integration of \glspl{LLM} in 6G networks (see Fig.~\ref{fig:LNF_challenges}), and proposes novel approaches along with promising research areas:
\smallskip

\textbf{Data Volumes and Quality.} \glspl{LLM} should process large volumes of heterogeneous data. \glspl{MNO} can harness knowledge distillation techniques \cite{Acharya2024Survey} to generate smaller focused datasets using domain-specific expert models, inherited from their conventional \gls{OAM}, to teach a general-purpose \gls{SOTA} \gls{LLM}. Particularly, this will extend the \gls{LLM}'s vocabulary with domain-relevant terminology, e.g. to understand specialized jargon used in trouble tickets. Efficient \gls{MNO}'s retention policies are also helpful to store, manage, and eventually delete internal data (e.g. network logs) within an optimal duration, weighting training needs and storage costs.

On the other hand, the performance of the \glspl{LLM} for \gls{MNO}-specific tasks is heavily dependent on the quality and breadth of the data. This calls for federated data governance policies that clean, compress and structure data locally at the data sources level before uploading to a central repository, provided that governance tools are perfectly aligned in multi-vendor environments to avoid conflicts.
\smallskip

\textbf{Computational Complexity.} The \gls{LLM} requires strong computational power and storage to analyze data effectively and deliver insights promptly. Multi-\gls{GPU} training leveraging parallel processing capabilities across multiple \glspl{GPU} can be used to distribute the training workload. Several approaches such as data parallelism, model parallelism, pipeline parallelism and hybrid parallelism can be envisioned to assign to each \gls{GPU}, a portion of the data, a slice of the model, a specific layer of the model, or a mix of the former blocks, respectively, to be processed in parallel. This introduces, however, an increased communication overhead and requires a seamless synchronization between the parallel \glspl{GPU}.

The \gls{LLM} may also be instantiated as distributed \glspl{NF} across edge, metro, and cloud segments. At the edge, \gls{LLM} instances can handle low-latency operations and localized decision-making such as tackling disruptions related to radio link. Metro \gls{LLM} instances can analyze region-wide trends, e.g. to deal with mobility-related anomalies across several cells. At the cloud, the focus is on centralized decision-making \cite{Chaparadza2023optimization} such as global outages. It is challenging to maintain a real-time orchestration and scale dynamically across these segments to meet fluctuating demands especially in heterogeneous hardware and software environments.

Additionally, computationally-intensive \gls{LLM} operations lead to significant carbon footprint calling for green energy initiatives, carbon offsetting and hosting datacenters in cooler climates.
\smallskip

\textbf{Security and Privacy Concerns.} \gls{LLM} suffers from adversarial attacks and data leakage vulnerabilities. The first attack relies on intentionally crafting the input data to corrupt the \gls{LLM} decisions. The training dataset should incorporate examples of such attacks to boost its resilience. The second vulnerability is designed to infer sensitive information from \gls{LLM} outputs. It can be mitigated by differential privacy that injects artificial randomness to data or computation process to protect data confidentiality. The amount of the injected noise should be carefully controlled to maintain the accuracy of the \gls{LLM} results.

Moreover, \glspl{MNO} tend to build their private \glspl{LLM} using open-source solutions (e.g. \gls{LLaMA}, DeepSeek) instead of commercial ones to avoid exposing sensitive data to third-party services \cite{Kan2024mobile_llama}.
\smallskip

\textbf{\gls{MNO} acceptance of \gls{LLM}-generated insights.} Fully autonomous \glspl{LLM} within 6G networks could potentially be perceived as "black boxes". Explainable \gls{AI} principles enable \glspl{LLM} to document their decision-making processes, making their actions understandable and auditable \cite{maatouk2024large}. Self-governance mechanisms allow these systems to self-regulate and mitigate errors, fostering confidence in their autonomy. The complexity of the \gls{LLM} models and the high-dimensionality of the data make it challenging to trace the decision-making process without reliance on approximations and simplifications.
\smallskip

\textbf{Lack of telecom-centric \gls{AI} standards.}
Due to the lack of \gls{AI} standardization efforts in the telecommunication field, each \gls{MNO} can take a distinct approach to embrace the \gls{LLM} technology. This may lead to adopting suboptimal \glspl{LLM} or disparate \gls{AI} practices, leading to incompatibilities. Initiatives such as ITU-T Y.3000 series and GSMA Responsible \gls{AI} Maturity Roadmap can orient telecommunication stakeholders in adopting \gls{AI} and \glspl{LLM} both strategically and ethically.




\section{Conclusions}
\label{sec:conclusion}
This paper investigates the architectural designs for making the \gls{LLM} a key functional element of 6G networks. The deployment form of the \gls{LLM} depends on the \gls{MNO}'s goals. The standalone design enjoys flexibility, sophisticated capabilities, and multi-system interoperability, whereas the fully integrated architecture promotes real-time performance, streamlined architecture, and coherence. The hybrid approach offers the best of both worlds, but remains technically complicated and costly to implement. An external \gls{LLM} poses serious security challenges, as opposed to the aforementioned architectures secured through \gls{SbD} principles. In their path towards integrating \glspl{LLM} in their ecosystem, \glspl{MNO} can begin with the standalone design and smoothly transition to the hybrid model over time through gradually upgrading an existing \gls{NF} such as \gls{NWDAF} to host the \gls{Near-RT} \gls{LLM}. This integration represents a transformative approach toward holistic intelligence and full automation promises of 6G.  A proof-of-concept based on the architectural framework proposed in this study is ongoing and will be a part of a follow-up work to this paper. 


\section*{Acknowledgement}

\begin{spacing}{}
    \bibliographystyle{IEEEtran}
    \bibliography{Main.bib}

\begin{thebibliography}{10}
\providecommand{\url}[1]{#1}
\csname url@samestyle\endcsname
\providecommand{\newblock}{\relax}
\providecommand{\bibinfo}[2]{#2}
\providecommand{\BIBentrySTDinterwordspacing}{\spaceskip=0pt\relax}
\providecommand{\BIBentryALTinterwordstretchfactor}{4}
\providecommand{\BIBentryALTinterwordspacing}{\spaceskip=\fontdimen2\font plus
\BIBentryALTinterwordstretchfactor\fontdimen3\font minus \fontdimen4\font\relax}
\providecommand{\BIBforeignlanguage}[2]{{%
\expandafter\ifx\csname l@#1\endcsname\relax
\typeout{** WARNING: IEEEtran.bst: No hyphenation pattern has been}%
\typeout{** loaded for the language `#1'. Using the pattern for}%
\typeout{** the default language instead.}%
\else
\language=\csname l@#1\endcsname
\fi
#2}}
\providecommand{\BIBdecl}{\relax}
\BIBdecl

\bibitem{chaoub2023hybrid}
A.~Chaoub, A.~Mämmelä, P.~Martinez-Julia, R.~Chaparadza, M.~Elkotob, L.~Ong, D.~Krishnaswamy, A.~Anttonen, and A.~Dutta, ``{Hybrid Self-Organizing Networks: Evolution, Standardization Trends, and a 6G Architecture Vision},'' \emph{IEEE Communications Standards Magazine}, vol.~7, no.~1, pp. 14--22, 2023.

\bibitem{vaswani2017attention}
A.~Vaswani, N.~Shazeer, N.~Parmar, J.~Uszkoreit, L.~Jones, A.~N. Gomez, L.~Kaiser, and I.~Polosukhin, ``{Attention is all you need},'' in \emph{Proceedings of the 31st International Conference on Neural Information Processing Systems}, 2017, p. 6000–6010.

\bibitem{maatouk2024large}
A.~Maatouk, N.~Piovesan, F.~Ayed, A.~De~Domenico, and M.~Debbah, ``{Large Language Models for Telecom: Forthcoming Impact on the Industry},'' \emph{IEEE Communications Magazine}, vol.~63, no.~1, pp. 62--68, 2025.

\bibitem{Kan2024mobile_llama}
K.~B. Kan, H.~Mun, G.~Cao, and Y.~Lee, ``{Mobile-LLaMA: Instruction Fine-Tuning Open-Source LLM for Network Analysis in 5G Networks},'' \emph{IEEE Network}, vol.~38, no.~5, pp. 76--83, 2024.

\bibitem{Wang2024NetConfEval}
C.~Wang, M.~Scazzariello, A.~Farshin, S.~Ferlin, D.~Kosti\'{c}, and M.~Chiesa, ``{NetConfEval: Can LLMs Facilitate Network Configuration?}'' \emph{Proc. ACM Netw.}, vol.~2, no. CoNEXT2, 2024.

\bibitem{Chen2023Knowledge}
Z.~Chen, W.~Zhang, Y.~Huang, M.~Chen, Y.~Geng, H.~Yu, Z.~Bi, Y.~Zhang, Z.~Yao, W.~Song, X.~Wu, Y.~Yang, M.~Chen, Z.~Lian, Y.~Li, L.~Cheng, and H.~Chen, ``{Tele-Knowledge Pre-training for Fault Analysis},'' in \emph{2023 IEEE 39th International Conference on Data Engineering (ICDE)}, 2023, pp. 3453--3466.

\bibitem{Said2024instruct}
A.~I.~A. Said, A.~Mekrache, K.~Boutiba, K.~Ramantas, A.~Ksentini, and M.~Rahmani, ``{5G INSTRUCT Forge: An Advanced Data Engineering Pipeline for Making LLMs Learn 5G},'' \emph{IEEE Transactions on Cognitive Communications and Networking}, pp. 1--1, 2024.

\bibitem{Bariah2024Large}
L.~Bariah, Q.~Zhao, H.~Zou, Y.~Tian, F.~Bader, and M.~Debbah, ``{Large Generative AI Models for Telecom: The Next Big Thing?}'' \emph{IEEE Communications Magazine}, vol.~62, no.~11, pp. 84--90, 2024.

\bibitem{Zhou2024survey}
H.~Zhou, C.~Hu, Y.~Yuan, Y.~Cui, Y.~Jin, C.~Chen, H.~Wu, D.~Yuan, L.~Jiang, D.~Wu, X.~Liu, C.~Zhang, X.~Wang, and J.~Liu, ``{Large Language Model (LLM) for Telecommunications: A Comprehensive Survey on Principles, Key Techniques, and Opportunities},'' \emph{IEEE Communications Surveys \& Tutorials}, pp. 1--1, 2024.

\bibitem{Khaloopour2024Resilience}
L.~Khaloopour, Y.~Su, F.~Raskob, T.~Meuser, R.~Bless, L.~Janzen, K.~Abedi, M.~Andjelkovic, H.~Chaari, P.~Chakraborty, M.~Kreutzer, M.~Hollick, T.~Strufe, N.~Franchi, and V.~Jamali, ``{Resilience-by-Design in 6G Networks: Literature Review and Novel Enabling Concepts},'' \emph{IEEE Access}, vol.~12, pp. 155\,666--155\,695, 2024.

\bibitem{Bariah2023understanding}
L.~Bariah, H.~Zou, Q.~Zhao, B.~Mouhouche, F.~Bader, and M.~Debbah, ``{Understanding Telecom Language Through Large Language Models},'' in \emph{GLOBECOM 2023 - 2023 IEEE Global Communications Conference}, 2023, pp. 6542--6547.

\bibitem{soman2023observations}
S.~Soman and R.~HG, ``{Observations on LLMs for telecom domain: capabilities and limitations},'' in \emph{Proceedings of the Third International Conference on AI-ML Systems}, 2023, pp. 1--5.

\bibitem{Chaparadza2023optimization}
R.~Chaparadza, A.~Chaoub, B.~Chng, N.~Davis, A.~Dutta, M.~Elkotob, D.~Krishnaswamy, K.~Mahdi, A.~Mämmelä, P.~Martinez-Julia, N.~K. Narang, L.~Ong, M.~Patwary, M.~Simsek, J.~Voigt, C.~Polk, K.~McDonnell, J.~Niemöller, D.~Milham, and J.~Cadman, ``{INGR Roadmap System Optimization Chapter},'' in \emph{2023 IEEE Future Networks World Forum (FNWF)}, 2023, pp. 1--89.

\bibitem{song2023restgpt}
Y.~Song, W.~Xiong, D.~Zhu, W.~Wu, H.~Qian, M.~Song, H.~Huang, C.~Li, K.~Wang, R.~Yao, Y.~Tian, and S.~Li, ``{RestGPT: Connecting Large Language Models with Real-World RESTful APIs},'' \emph{arXiv preprint arXiv:2306.06624}, 2023.

\bibitem{Acharya2024Survey}
K.~Acharya, A.~Velasquez, and H.~H. Song, ``{A Survey on Symbolic Knowledge Distillation of Large Language Models},'' \emph{IEEE Transactions on Artificial Intelligence}, vol.~5, no.~12, pp. 5928--5948, 2024.

\end{thebibliography}
\end{spacing}
\section*{Biography}
\label{sec:bio}
\small
\begin{IEEEbiographynophoto}{Abdelaali Chaoub} [SM] has been an Associate Professor in Telecommunications, appointed at the Institut National des Postes et Télécommunications (INPT) of Morocco since 2015. He holds an engineering degree in Telecommunication from INPT in 2007 with the highest honors, and a Ph.D. degree in Telecommunication from Mohammed V-Agdal University in 2013. His research interests are in the area of 6G wireless communications and networks, including architectural contributions to autonomous systems, remote and rural connectivity solutions, design and optimization of IoT-enabled smart environments, cognitive radio, and multimedia communications. He is a paper reviewer for several leading international journals and conferences. He has accumulated intersectoral skills through work experience both in academia and industry as a senior VoIP Consultant then as a Lead Standardization Specialist (2007 – 2015).
\end{IEEEbiographynophoto}

\begin{IEEEbiographynophoto}{Muslim Elkotob} [M] Dr.-Ing./PhD, is a Principal Solutions Architect at Vodafone with a lead role and end-to-end responsibility in the Enterprise Business Line. He works on driving innovation and standardizing architectures in SDN/NFV, Autonomics, Slicing and Security areas in 5G and IoT. An IPv6-Forum Fellow and delegate with lead roles in various SDOs including ETSI, TMForum, ITU-T and IEEE. Having a career background with vendors, service providers and R\&D, he has spent the last seven years strengthening Vodafone's role in the enterprise Value Chain as a global player with a powerful infrastructure and autonomic IT services on top.
\end{IEEEbiographynophoto}%

\end{document}